\documentclass[lettersize,journal]{IEEEtran}
\usepackage{amsmath,amsfonts}
\usepackage{algorithmic}
\usepackage{algorithm}
\usepackage{array}
\usepackage[caption=false,font=footnotesize]{subfig}
\usepackage{textcomp}
\usepackage{stfloats}
\usepackage{xcolor}

\usepackage{url}
\usepackage{verbatim}
\usepackage{graphicx}
\usepackage{cite}
\usepackage{orcidlink}
\hypersetup{hidelinks} 
\usepackage{hyperref}
\hypersetup{
    colorlinks=true,
    linkcolor=black,
    citecolor=black,
    urlcolor=black
}
\usepackage{enumitem}
\setlist[itemize]{topsep=2pt,parsep=2pt,itemsep=2pt,partopsep=2pt}
\setlist[itemize]{leftmargin=10pt}

\begin{document}

\title{\vspace{-0.5cm} An Experimental Study on Fine-Grained Bistatic Sensing of UAV Trajectory via Cellular Downlink Signals \vspace{-0.2cm}}

\author{\IEEEauthorblockN{Chenqing Ji, Jiahong Liu, Qionghui Liu, Yifei Sun, Chao Yu and Rui Wang}
\thanks{This work was supported in part by the National Natural Science Foundation of China (NSFC) under Grant U25A20389, in part by Shenzhen Science and Technology Program under Grant JCYJ20241202125328038, and in part by High Level of Special Funds under Grant G03034K004. \textit{(Corresponding author: Rui Wang)}

The authors are with the Department of Electronic and Electrical Engineering, Southern University of Science and Technology, Shenzhen 518055, China (e-mail: \{12332152,12210631,12212010,sunyf2019,12431241\}@mail.sustech.edu.cn; wangr@sustech.edu.cn).}
\vspace{-1.2cm}}

\markboth{}%
{Shell \MakeLowercase{\textit{et al.}}: A Sample Article Using IEEEtran.cls for IEEE Journals}

\maketitle
\begin{abstract}
In this letter, a dual-bistatic unmanned aerial vehicles (UAVs) tracking system utilizing downlink Long-Term Evolution (LTE) signals is proposed and demonstrated. Particularly, two LTE base stations (BSs) are exploited as illumination sources. Two passive sensing receivers are deployed at different locations to detect the bistatic Doppler frequencies of the target UAV at different directions according to downlink signals transmitted from their corresponding BSs, such that the velocities of the UAV versus time can be estimated. Hence, the trajectories of the target UAV can be reconstructed. Although both the target UAV and the sensing receivers are  around $\mathbf{200}$ meters away from the illuminating BSs, it is demonstrated by experiments that the tracking errors are below $\mathbf{50}$ centimeters for $\mathbf{90\%}$ of the complicated trajectories, when the distances between the UAV and sensing receivers are less than $\mathbf{30}$ meters. Note this accuracy is significantly better than the ranging resolution of LTE signals, high-accuracy trajectory tracking for UAV might be feasible via multi-angle bistatic Doppler measurements if the receivers are deployed with a sufficient density.
\end{abstract}
\vspace{-0.2cm}
\vspace{-0.5cm}
\section{Introduction}
With the rapid growth of the low-altitude economy, the number of UAVs has increased exponentially, which may raise safety risks especially in crowded urban area. Hence, the real-time monitoring of low-altitude UAVs has gained considerable significance.
The integrated sensing and communication (ISAC) technology has a great potential in monitoring low-altitude UAVs due to the widely deployed communication base stations (BSs). Upgrading the existing BSs with UAV sensing capability is challenging in signal design, resource allocation, interference management, and deployment costs\cite{10273396,10872780}. Therefore, utilizing existing commercial BSs as illumination sources for passive sensing is an attractive solution.

Although the performance of passive sensing has been investigated extensively \cite{6620961,8489883,7488178,8809814}, 
there are a few works focusing on the application of UAVs detection via downlink cellular signals. For example, a passive radar system utilizing the signals of Global System for Mobile Communications (GSM) was demonstrated in \cite{7411917}. It was shown that the Doppler frequency of the GSM signal scattered off a moving UAV with low radar cross-section (RCS) could be detected at a sensing receiver by correlating the scattered signal
with the line-of-sight (LoS) GSM signal. Furthermore, the first experimental evidence of passive UAV detection via commercial GSM BSs was provided in \cite{7497375}.
In \cite{https://doi.org/10.1049/joe.2019.0583} and \cite{10172437}, it was shown that the Long-Term Evolution (LTE) and the fifth-generation (5G) mobile communication signals could be utilized to detect a low-RCS UAV, respectively. Recently, it was demonstrated in \cite{10902137} that by exploiting the pilot symbols of LTE signals, a low-RCS UAV can be  effectively detected in complex electromagnetic environments via the bistatic range,  Doppler and angle-of-arrival (AoA) information.
However, these works focused only on existence detection of UAVs, instead of localization or tracking. 

In fact, integrating the bistatic range, Doppler and AoA measurements of single or multiple passive receivers, it is feasible to track the trajectories of UAVs. For example in \cite{10.1145/3507657.3529658}, a small commercial UAV at a distance of $20$m could be localized via the bistatic ranges between multiple transmitters and a single receiver, i.e., multistatic sensing. In \cite{9764210}, a sensor fusion technique was proposed to integrate the bistatic range, Doppler and AoA information of GSM and LTE signals with cameras in UAV tracking. 
Furthermore, the LIPASE system proposed in \cite{sun2024experimentalstudypassiveuav} used a single sensing receiver to reconstruct the trajectory of a UAV based on the bistatic Doppler, AoA, and range information. However, the ranging resolution is limited by the signal bandwidth, and the accuracy of AoA estimation is constrained by the array size \cite{griffiths2022introduction,s23073435}.
Thus, either range or AoA information may hardly be sufficient in high-accuracy trajectory tracking.
In contrast to range and AoA measurements, improving the detection accuracy of the bistatic Doppler frequency is much easier, as it only requires longer signal duration in cross ambiguity function (CAF) calculation \cite{6915995,8666692}.

Hence, integrating the bistatic Doppler frequency measurements of multiple directions, we would like to show in this letter that high-accuracy UAV trajectory tracking via cellular signals may be feasible. 
Specifically, the proposed tracking system consists of two passive sensing receivers. Each receiver is equipped with two radio frequency (RF) chains to detect the bistatic Doppler frequency of the target UAV. Hence, the velocities and trajectories of the target UAV can be reconstructed  with the knowledge of initial UAV location. 
It is demonstrated that $90\%$ of the trajectory tracking errors are below $0.5$m, when the UAV-receiver distances are within $30$m and the BS-UAV distances are larger than $200$m.
$90\%$ of the trajectory tracking errors degrade to $1.2$m when the estimation error on initial UAV location is considered. To the best of our knowledge, this is the first work demonstrating the passive UAV tracking with multiple sensing receivers and Doppler-only measurement.

\vspace{-0.4cm}
\section{System Overview} \label{sec2}
The proposed passive sensing system for UAV trajectory tracking consists of one or two LTE eNBs and two receivers. If there is a single eNB, it is used as the illumination source of both receivers. 
If two eNBs are involved, they serve as the illumination sources of the two receivers, respectively. We only consider the latter dual-bistatic case for the elaboration convenience, however, the tracking approaches are similar.
Due to the frequency reuse, the two eNBs are working on two distinct frequency bands. The two eNBs and two receivers are referred to as the BS 1, BS 2, Receiver 1 and Receiver 2, respectively. The Receiver 1 and 2 are receiving signals on the frequency bands of BS 1 and 2, respectively. 

\begin{figure}[!ht]
    \centering
    \vspace{-0.46cm}
    \includegraphics[width=0.43\textwidth]{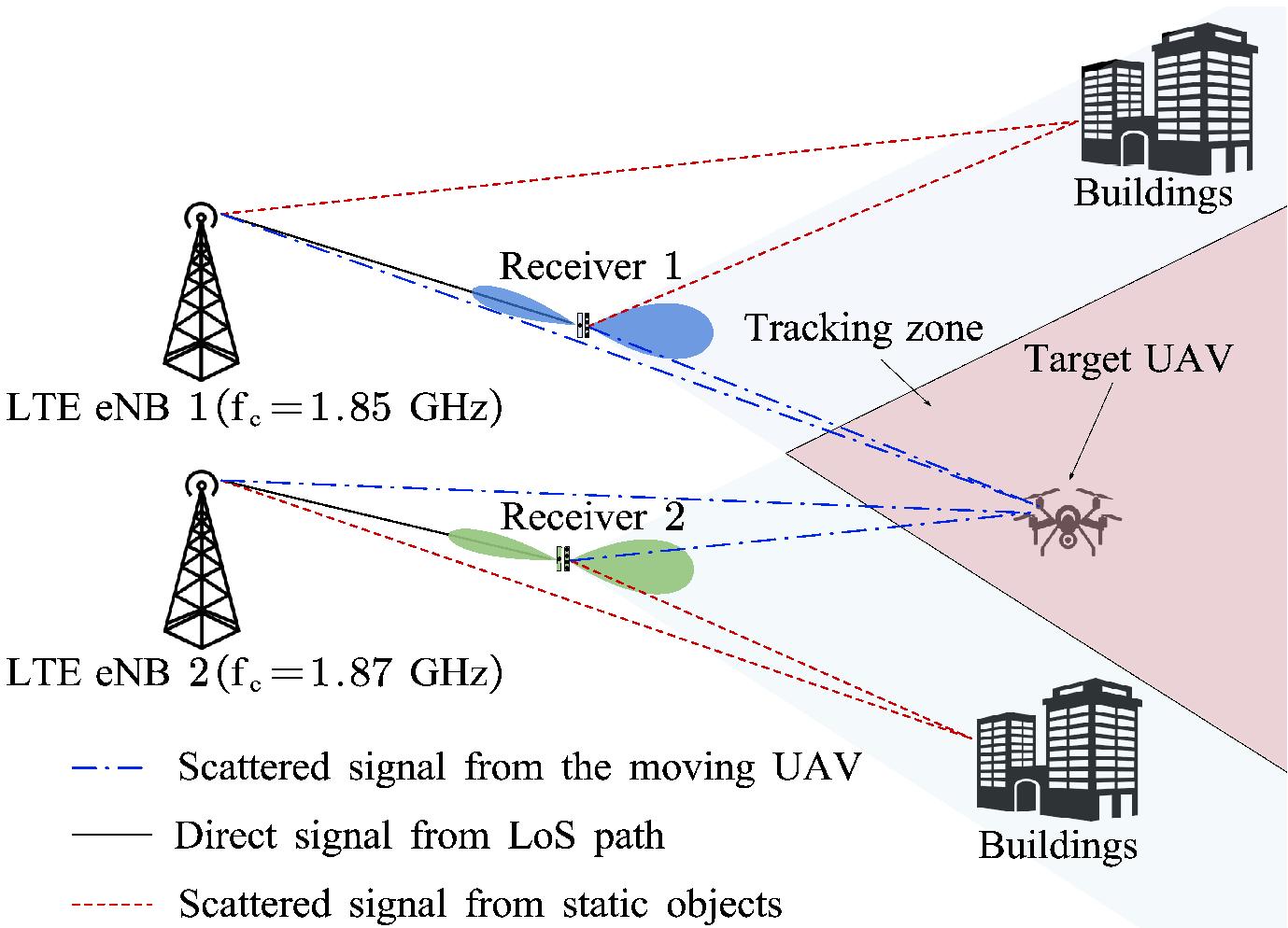}
     \vspace{0.1cm}
    \caption{An example scenario of the proposed passive UAV tracking, where the carrier frequencies of the BSs are the measured ones in the experiment.}
    \label{fig1}
    \vspace{-0.45cm}
\end{figure}

Both receivers are deployed to measure the bistatic Doppler frequencies of one target UAV. Particularly, there are two RF chains at each receiver.
One narrow receive beam is aligned with the LoS direction of the corresponding BS, and the other wide receive beam is targeted at the area of trajectory tracking, as shown in Fig. \ref{fig1}.
The LoS path between the BS $i$ ($i\!=\!1,2$) and the Receiver $i$ is referred to as the $i$-th reference channel, and the corresponding BS-UAV-receiver path is referred to as the $i$-th surveillance channel. By comparing the received signals of both channels, each receiver is able to detect the bistatic Doppler frequency of the target UAV at one direction. The initial location of the UAV can be estimated via the AoAs detected at both receivers, and the two-dimensional velocity of the UAV can be reconstructed for trajectory tracking.
As a remark, the proposed method can be extended to three-dimensional tracking if more receivers are deployed. 

\vspace{-0.2cm}
\section{Bistatic Doppler frequency detection} \label{sec3}
\subsection{Signal Model}
Let $s_{i}(t)$, $ t\in [0, \mathrm{T}]$, be the information-bearing signal transmitted by the BS $i$ ($i\!=\!1,2$). The received signal of the reference channel at the Receiver $i$, denoted as $y_{r,i}(t)$, can be expressed as
\vspace{-0.2cm}
\begin{equation}
y_{r,i}(t)=\alpha_{r,i}s_{i}\left(t-\tau_{r,i}\right)+n_{r,i}(t),
\vspace{-0.1cm}
\label{eq1}
\end{equation}
where $\alpha_{r,i}$ and $\tau_{r,i}$ denote the complex gain and delay of the LoS path respectively, and $n_{r,i}(t)$ denotes the noise plus the interference in the reference channel. The interference mainly comes from the signals scattered off the surrounding scattering clusters. 
Note that, due to the loss of scattering, the power of the interference and noise is usually negligible compared with the strong signal power from the LoS path.

Meanwhile, the received signal of the surveillance channel at the Receiver $i$, denoted as $y_{s,i}(t)$, consists of the echo signals scattered off the target UAV and surrounding static scattering clusters. It can be expressed as
\vspace{-0.1cm}
\begin{equation}
\begin{aligned}
y_{s,i}(t) & =\alpha_{s,i}^{\mathrm{tar}}(t)s_{i}\left(t-\tau_{s,i}^{\mathrm{tar}}(t)\right)e^{j2\pi f_{i}^{\mathrm{D,tar}}(t)t} \\
 & +\sum_{l=0}^{L_{s,i}}\alpha_{s,i}^ls_{i}\left(t-\tau_{s,i}^l\right)+n_{s,i}(t), \label{eq2}
 \end{aligned}
\vspace{-0.1cm}
\end{equation}
where $\alpha_{s,i}^{\mathrm{tar}}$, $\tau_{s,i}^{\mathrm{tar}}(t)$ and $f_{i}^{\mathrm{D,tar}}(t)$ denote the 
complex gain, delay and Doppler frequency of the 
signal scattered off the UAV respectively, $L_{s,i}$ is the number of
signal paths from undesired static scattering clusters ($l =0$ refers to the undesired LoS path), $\alpha_{s,i}^l$, $\tau_{s,i}^l$ denote the complex gain and delay of the $l$-th undesired path respectively, and $n_{s,i}(t)$ denotes the noise.

The received signals of the reference channel and the surveillance channel at both the receivers are sampled with a period $\mathrm{T_s}$ for baseband processing, which can be written as $y_{r,i}[n]=y_{r,i}(n\mathrm{T}_\mathrm{s})$
and $y_{s,i}[n]=y_{s,i}(n\mathrm{T}_\mathrm{s})$, respectively, where $n = 0, 1, ..., \mathrm{T}/\mathrm{T_s}$ is the sample index.
Note that the signal components with zero Doppler frequency in $y_{s,i}[n]$ 
will raise significant interference to the estimation of the desired Doppler frequency. The multi-coherent integration time (Multi-CIT) based interference cancellation technique in \cite{9393557} is applied. Finally, the signal of the surveillance channel after the above interference cancellation at the Receiver $i$ is denoted as $\tilde{y}_{s,i}[n]$ ($i\!=\!1,2$).
\vspace{-0.3cm}
\subsection{Bistatic Doppler Frequency Detection} \label{sec3B}
In this part, the detection of time-varying bistatic Doppler frequency of the target UAV at each receiver is elaborated. It is assumed that the velocity of the UAV can be treated as a constant in a duration of $\mathrm{T_d}$ seconds. Hence, the CAF is calculated every $\mathrm{T_d}$ seconds at each receiver to detect the bistatic Doppler frequency. Due to the interference in the surveillance channel, an adaptive threshold mechanism is adopted to capture the bistatic Doppler frequency from CAF. Finally, a filtering algorithm is proposed to suppress the false alarm and miss detection in the Doppler detection.

Particularly, the time instances $t=k\mathrm{T_d}$ ($k=0,1,2,...$), when the bistatic Doppler frequency is detected, are referred to as the detection time instances. At each detection time instance, a window of $N_{w}$ baseband samples (thus $N_{w}\mathrm{T_s}$ seconds) is utilized to calculate the CAF. The CAF between the received
signals of the reference channel and
the surveillance channel at the Receiver $i$ and the $k$-th detection time instance can be expressed as
\vspace{-0.1cm}
\begin{equation}
R_{i,k}(f_D)=\max_{\tau_{i,k}}\sum_{n=kN_0}^{kN_0+N_w-1}\tilde{y}_{s,i}[n]y_{r,i}^*[
n-\tau_{i,k}]e^{-j2\pi f_Dn\mathrm{T}_\mathrm{s}},
\vspace{-0.1cm}
\end{equation}
where $N_{0} = \mathrm{T_{d}/T_{s}}$, and $(.)^*$ is the complex conjugate. Note that due to the limited bandwidth of communication signals, the propagation delay between the received signals from the reference channel and the surveillance channel, denoted as $\tau_{i,k}$, can be neglected. 

Without interference and noise, $R_{i,k}(f_D)$ usually has a single peak at $f_D = f_{i}^\mathrm{D,tar}{(k\mathrm{T}_\mathrm{d})}$.
In practice, due to noise and interference (e.g., the micro-Doppler effects from the rotation of the UAV propellers), it might be possible to detect multiple peak values at one detection time instance from $R_{i,k}(f_D)$. To suppress these false alarms, an adaptive threshold-based filtering technique is utilized, which adaptively changes the detection threshold of Doppler frequency according to the local background noise level.
Particularly, a bistatic Doppler frequency $f_{D}$ is detected at the Receiver $i$ and the $k$-th detection time instance if $\left|R_{i,k}(f_D)\right|\geq\beta_{i,k}(f_D).$ The detection threshold $\beta_{i,k}(f_D)$ is defined as
\begin{equation}
\beta_{i,k}(f_{D})=\frac{\gamma}{2P+1}\sum_{p=-P}^{P}\big|R_{i,k}(f_{D}+p\Delta f)\big|,
\vspace{-0.1cm}
\end{equation}
where $\gamma > 1$ is a
scaling factor, $P$ is a constant, $\Delta f=\frac{1}{N_w\mathrm{T_s}}$ is the detection resolution of the bistatic Doppler frequency. Thus, a bistatic Doppler frequency $f_{D}$ is detected when the corresponding CAF magnitude is significantly greater than those of its neighboring frequencies.

As a remark, although the above adaptive method could effectively suppress the majority of false alarms in Doppler detection, an excessive threshold in the low-SNR condition at some detection time instances may discard all the peak values (i.e., miss detection), and an overly low threshold at some detection time instances may still leave residual false alarms. Thus, the issue of miss detection and false alarm may still exist. Hence, it is necessary to represent the detected bistatic Doppler frequency at the Receiver $i$ and the $k$-th detection time instance by the following set:
\vspace{-0.1cm}
\begin{equation}
    \mathcal{D}_{i,k} =\big\{f_{i,k,m}\big||R_{i,k}(f_{i,k,m})|\geq\beta_{i,k}(f_{i,k,m}),\forall m>0\big\}, 
\vspace{-0.1cm}
\end{equation}
where $f_{i,k,m}$ is the $m$-th detected bistatic Doppler frequency. $\mathcal{D}_{i,k}$ can be empty, or with single or multiple elements.
 
To reconstruct the trajectory of UAV, a single bistatic Doppler frequency for each receiver is requested at one detection time instance. Hence, the following algorithm with interpolation and weighted averaging can be adopted to handle the issue of miss detection and false alarm:
\begin{itemize}
    \item Step 1: Find the first detection time instance with single detected Doppler frequency, say the $k$-th detection time instance. Let $\hat{f}_{i,k}=f_{i,k,1}$ ($i\!=\!1,2$) and $j=k+1$.
    \item Step 2: If  $|\mathcal{D}_{i,j}|=1$ ($i\!=\!1,2$), the estimation of Doppler frequency is given by $\hat{f}_{i,j}=f_{i,j,1}$. Otherwise, if $|\mathcal{D}_{i,j}|>1$, the estimation of Doppler frequency can be determined by combining the previous detection with the current detections via weighted averaging as
    \vspace{-0.3cm}
    \begin{equation}
    \mbox{\bf False Alarm: }\hat{f}_{i,j} = \alpha \cdot \hat{f}_{i,j-1} + (1-\alpha) \sum_{m=1}^{\vert\mathcal{D}_{i,j}\vert} p_{i,j,m}\cdot f_{i,j,m}, \nonumber
    \vspace{-0.1cm}
    \end{equation}
    where $p_{i,j,m}$ and $\alpha \in [0, 1]$ are both weights. If $|\mathcal{D}_{i,j}|=0$, the linear interpolation is utilized to fill this miss detection based on neighboring valid detections as
    \vspace{-0.1cm}   
    \begin{equation}
    \mbox{\bf Miss detection: }\hat{f}_{i,j} = \hat{f}_{i,j-1} + \frac{\hat{f}_{i,l}-\hat{f}_{i,j-1}}{l-(j-1)}, \nonumber
    \vspace{-0.1cm}
    \end{equation}
    where $l=\min\{m|\vert\mathcal{D}_{i,m}\vert=1,m>j\} $.
    \item Step 3: The algorithm terminates at the last detection time instance. Otherwise, $j=j+1$ and go to Step 2.
\end{itemize}

Based on the above algorithm, the estimations of bistatic Doppler frequencies at the two receivers, $\{\hat{f}_{1,j},\hat{f}_{2,j}|\forall j\}$, can be uniquely determined. Finally, the Kalman filter is adopted to smooth the estimations. The estimations of bistatic Doppler frequencies at the Receiver $i$ ($i\!=\!1,2$) after Kalman filter are denoted as $\{\tilde{f}_{i,j}|\forall j\}$. In fact, according to our experiment, there is single Doppler detection in most of the time instances. In the following elaboration, it is assumed that $\tilde{f}_{i,j}$ ($i\!=\!1,2$) can be obtained for all the detection time instances. As a remark, even with sudden LoS blockages, the bistatic Doppler detection still maintains the robustness based on the cumulative effect of signal energy within the CAF time window. Moreover, if miss detection is caused by LoS blockage, the above algorithm may still recover the bistatic Doppler frequency via interpolation.
\vspace{-0.2cm}

\section{UAV Trajectory Tracking} \label{sec4}
In this section, the UAV trajectory tracking method is elaborated given the smoothed bistatic Doppler frequency estimations $\{\tilde{f}_{i,k}| \forall i,k\}$ at the two receivers. The location vectors of two LTE BSs and
two receivers are denoted as $\mathbf{p}_{1}^{tx} = [x_{\mathrm{T1}}, y_{\mathrm{T1}}]^{\mathrm{T}}$, $\mathbf{p}_{2}^{tx} = [x_{\mathrm{T2}}, y_{\mathrm{T2}}]^{\mathrm{T}}$, $\mathbf{p}_{1}^{rx} = [0, 0]^{\mathrm{T}}$ and $\mathbf{p}_{2}^{rx} = [x_{\mathrm{R2}}, y_{\mathrm{R2}}]^{\mathrm{T}}$,
respectively.
They can be measured in advance with high accuracy.

Let $\mathbf{p}_{k} = [x_{k}, y_{k}]^{\mathrm{T}}$ and $\mathbf{v}_{k} = [v_{xk}, v_{yk}]^{\mathrm{T}}$ be the location and velocity of the target UAV in a two-dimensional plane at the $k$-th detection time instance, we have 
\vspace{-0.2cm}
\begin{equation}
\mathbf{p}_{k+1} = \mathbf{p}_{k} + \mathbf{v}_{k} \mathrm{T_d}. 
\vspace{-0.1cm}
\end{equation}
Let $f_{1,k}$ and $f_{2,k}$ denote the true bistatic Doppler frequencies of the UAV at the $k$-th detection time instance and the Receiver 1 and 2, respectively. We have
\vspace{-0.2cm}
\begin{equation}
\begin{bmatrix}
f_{1,k} \\
f_{2,k}
\end{bmatrix}=\mathbf{D}_k\cdot\mathbf{v}_k,
\end{equation}
or 
\vspace{-0.4cm}
\begin{equation}
\mathbf{v}_k=\mathbf{D}_k^{-1}\cdot
\begin{bmatrix}
f_{1,k} \\
f_{2,k}
\end{bmatrix}, \label{eq7}
\vspace{-0.1cm}
\end{equation}
where 
\vspace{-0.5cm}
\begin{equation}
\mathbf{D}_k = -
\begin{bmatrix}
    \frac{1}{\lambda_1} \left( \frac{\mathbf{p}_{k} - \mathbf{p}_{1}^{tx}}{\|\mathbf{p}_{k} - \mathbf{p}_{1}^{tx}\|} + \frac{\mathbf{p}_{k} - \mathbf{p}_{1}^{rx}}{\|\mathbf{p}_{k} - \mathbf{p}_{1}^{rx}\|} \right)^{\mathrm{T}} \\
    \frac{1}{\lambda_2} \left( \frac{\mathbf{p}_{k} - \mathbf{p}_{2}^{tx}}{\|\mathbf{p}_{k} - \mathbf{p}_{2}^{tx}\|} + \frac{\mathbf{p}_{k} - \mathbf{p}_{2}^{rx}}{\|\mathbf{p}_{k} - \mathbf{p}_{2}^{rx}\|} \right)^{\mathrm{T}}
    \end{bmatrix},
\vspace{-0.1cm}
\end{equation}
and $\lambda_1$ and $\lambda_2$ denote the wavelength of the illuminating signals from the BS 1 and 2 respectively. 

In this letter, it is assumed that the initial location of the target UAV $\tilde{\mathbf{p}}_{0} = [\tilde{x}_{0}, \tilde{y}_{0}]^{\mathrm{T}}$ has already been detected. Substituting the true bistatic Doppler frequencies $f_{1,k}$ and $f_{2,k}$ by their estimations $\tilde{f}_{1,k}$ and $\tilde{f}_{2,k}$ respectively, the trajectory of the target UAV can be tracked iteratively as follows:
\vspace{-0.1cm}
\begin{equation}
\tilde{\mathbf{p}}_{k+1} = \tilde{\mathbf{p}}_{k} + \tilde{\mathbf{v}}_{k} \mathrm{T_d}, k=0,1,...,
\vspace{-0.1cm}
\end{equation}
where 
\vspace{-0.3cm}
\begin{equation}
    \tilde{\mathbf{v}}_{k}= \tilde{\mathbf{D}}_{k}^{-1}\cdot
\begin{bmatrix}
\tilde{f}_{1,k} \\
\tilde{f}_{2,k}
\end{bmatrix}, 
\vspace{-0.1cm}
\end{equation}
and 
\vspace{-0.3cm}
\begin{equation}
    \tilde{\mathbf{D}}_{k}= -
\begin{bmatrix}
    \frac{1}{\lambda_1} \left( \frac{\tilde{\mathbf{p}}_{k} - \mathbf{p}_{1}^{tx}}{\|\tilde{\mathbf{p}}_{k} - \mathbf{p}_{1}^{tx}\|} + \frac{\tilde{\mathbf{p}}_{k} - \mathbf{p}_{1}^{rx}}{\|\tilde{\mathbf{p}}_{k} - \mathbf{p}_{1}^{rx}\|} \right)^{\mathrm{T}} \\
    \frac{1}{\lambda_2} \left( \frac{\tilde{\mathbf{p}}_{k} - \mathbf{p}_{2}^{tx}}{\|\tilde{\mathbf{p}}_{k} - \mathbf{p}_{2}^{tx}\|} + \frac{\tilde{\mathbf{p}}_{k} - \mathbf{p}_{2}^{rx}}{\|\tilde{\mathbf{p}}_{k} - \mathbf{p}_{2}^{rx}\|} \right)^{\mathrm{T}}
    \end{bmatrix}.
\vspace{-0.1cm}
\end{equation}

\begin{figure}[!t]
    \centering
    \vspace{-0.47cm}
    \subfloat[The overview of the stadium for experiment]{
        \includegraphics[width=0.32\textwidth,height=0.23\textheight
        ]{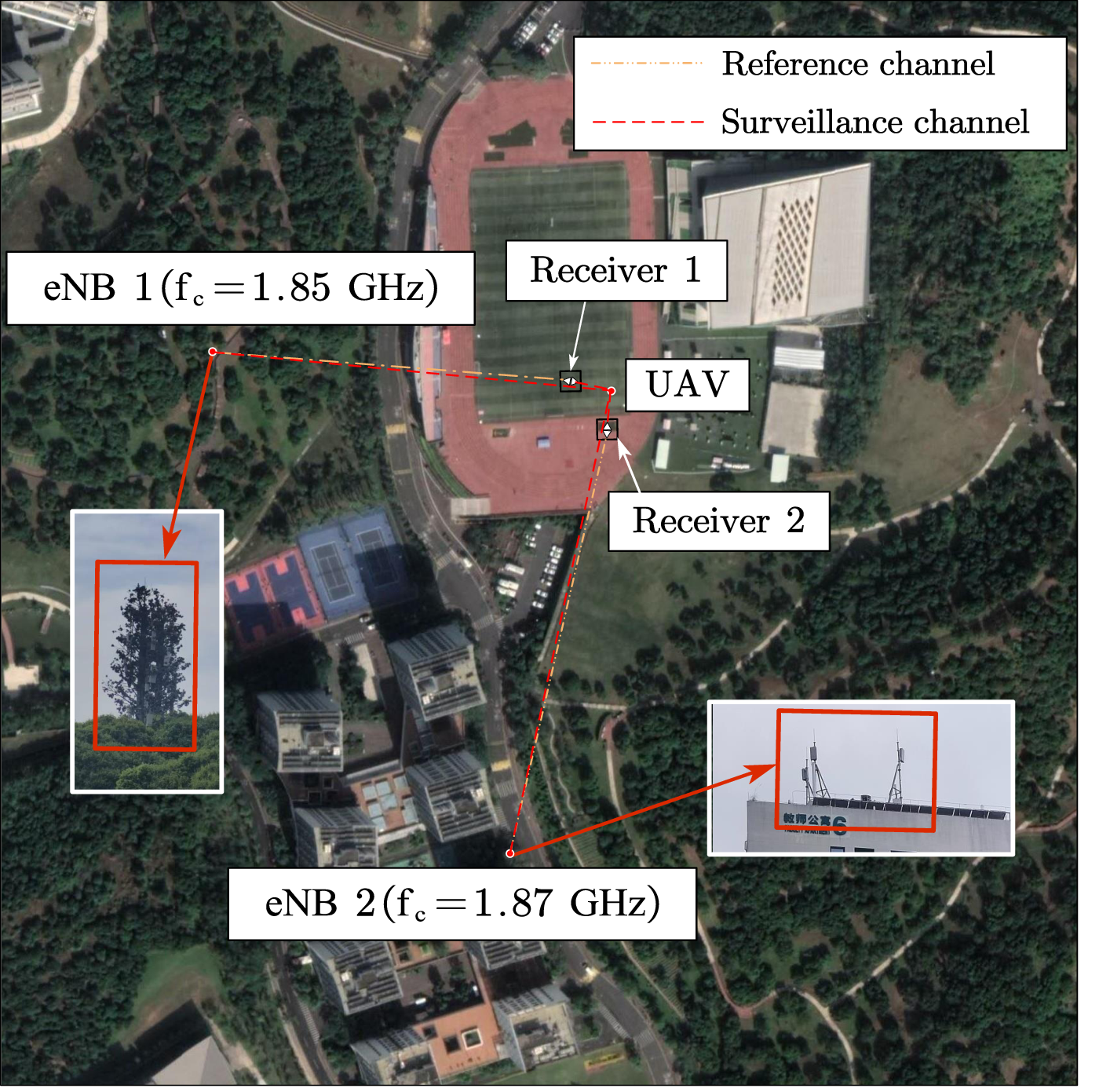}
    \label{fig4-a}}
    \hfill
    \subfloat[A snapshot of experiment scenario]{
        \includegraphics[width=0.35\textwidth]{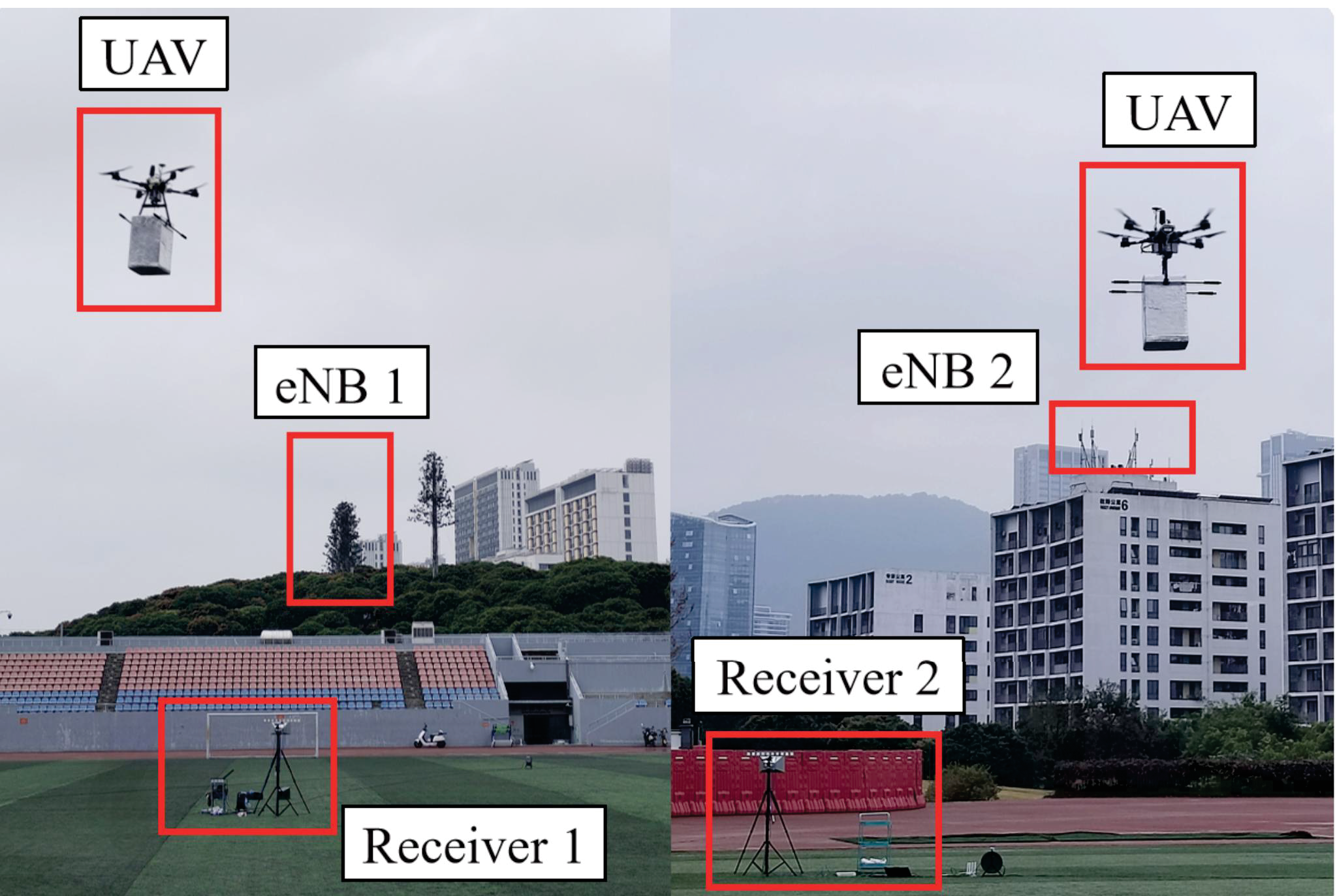}
    \label{fig4-b}}
  \caption{Illustration of experiment scenario.}
 \vspace{-0.62cm}
\end{figure}

\vspace{-0.2cm}
\section{Experiments and Discussion} \label{sec5}
To demonstrate the performance of the proposed passive sensing system, experiments on the trajectory tracking are conducted in a stadium, as illustrated in Fig. \ref{fig4-a}. The two sensing receivers are implemented by two devices of NI USRP-2953R. Particularly, each receiver consists of one horn antenna, one linear array, one USRP-2953R and one laptop. The horn antenna and linear array are used for reference and surveillance channels, respectively. The received signals of both channels are sampled at the USRP and processed by the laptop. In the processing of received signals, $N_{w}\mathrm{T_{s}}=0.5$s and 
$\mathrm{T_{d}}=0.05$s. Hence, the resolution of the bistatic Doppler frequency is
$1/N_{w}\mathrm{T_{s}} =2$Hz, and the resolution of bistatic range rate is around $0.16$m/s.

The experiment scenario is illustrated in Fig. \ref{fig4-b}. Two receivers are deployed on the soccer field, 
while the two LTE eNBs are at separate locations. One is on a hill and the other one is on the rooftop of a building. There are also buildings nearby. The scenario is close to the rooftop scenario in dense urban area, which is one of the application scenarios. The distances between the eNBs and the corresponding receivers are approximately $200$m and $220$m, respectively. The carrier frequencies of the two eNBs are $1.85$GHz and $1.87$GHz respectively, both are in frequency-division duplex (FDD) mode with a bandwidth of $20$MHz.

A quadcopter suspended with a $25\text{cm} \times 20\text{cm} \times 40\text{cm}$ cuboid cardboard box, which is wrapped in aluminum foil, is used as the target UAV in our experiment to imitate a delivery drone. The distances between the UAV and sensing receivers are around $30$m. Note that the BS-UAV distances are more than $200$m, increasing the UAV-receiver distances may decrease the BS-UAV distances. Hence, the received signal power may not significantly degrade.

Two flight trajectories are evaluated: one is U-shaped, and the other is a triangle.
The UAV is equipped with a RTK positioning module to obtain the ground truth of the trajectories with a centimeter-level accuracy. At the sensing receiver, a baseband signal of $2.5$MHz bandwidth is filtered from the $20$MHz downlink signal for Doppler frequency detection. Note that this baseband bandwidth is not sufficient for high-accuracy ranging. The dataset of raw received signals after sampling has been accessible online\footnote{\href{https://lasso525.quickconnect.cn/d/s/12gRTWCXynuW6Srov7kWhZRYhru1LXts/9QsN_gaKp7aB1_PTxVsC76w9JUuBhOLb-ebfgm6_tJgw}{
https://lasso525.quickconnect.cn}.}, where synchronous randomization is conducted to avoid the potential privacy leakage. 

\begin{figure}[!ht]
    \centering
    \vspace{-0.8cm}
    \subfloat[Receiver 1]{
        \includegraphics[width=0.23\textwidth]{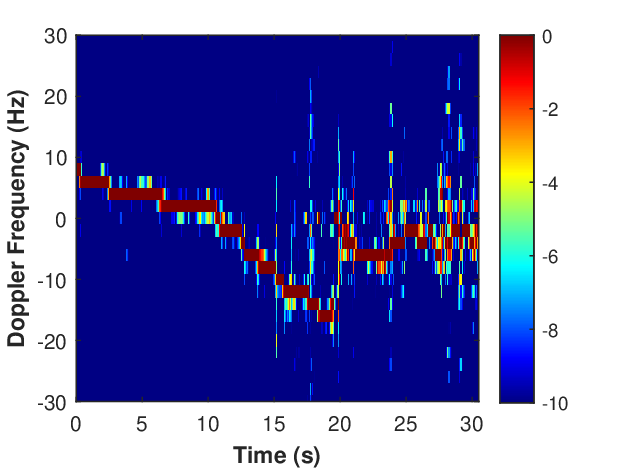}
    }
    \subfloat[Receiver 2]{
        \includegraphics[width=0.23\textwidth]{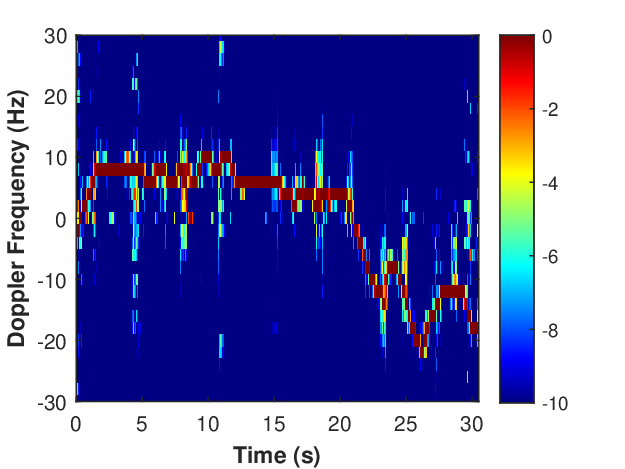}%
    }
    \caption{The time-Doppler spectrograms at the two receivers.}
    \label{fig5}
    \vspace{-0.3cm}
\end{figure}

The bistatic Doppler frequencies 
versus time (i.e., CAFs at different detection time instances) at the two receivers are illustrated in Fig. \ref{fig5}, where the target UAV flies along the U-shaped trajectory. It can be observed that the bistatic Doppler frequencies  
exhibit distinct patterns at the two receivers, due to their different surveillance channels.
The signs of the bistatic Doppler frequencies depend on the motion direction of the target UAV, indicating whether it is approaching or moving away from the corresponding receivers.

\begin{figure}[!htbp]
    \centering
    \vspace{-0.4cm}
    \includegraphics[width=0.44\textwidth]{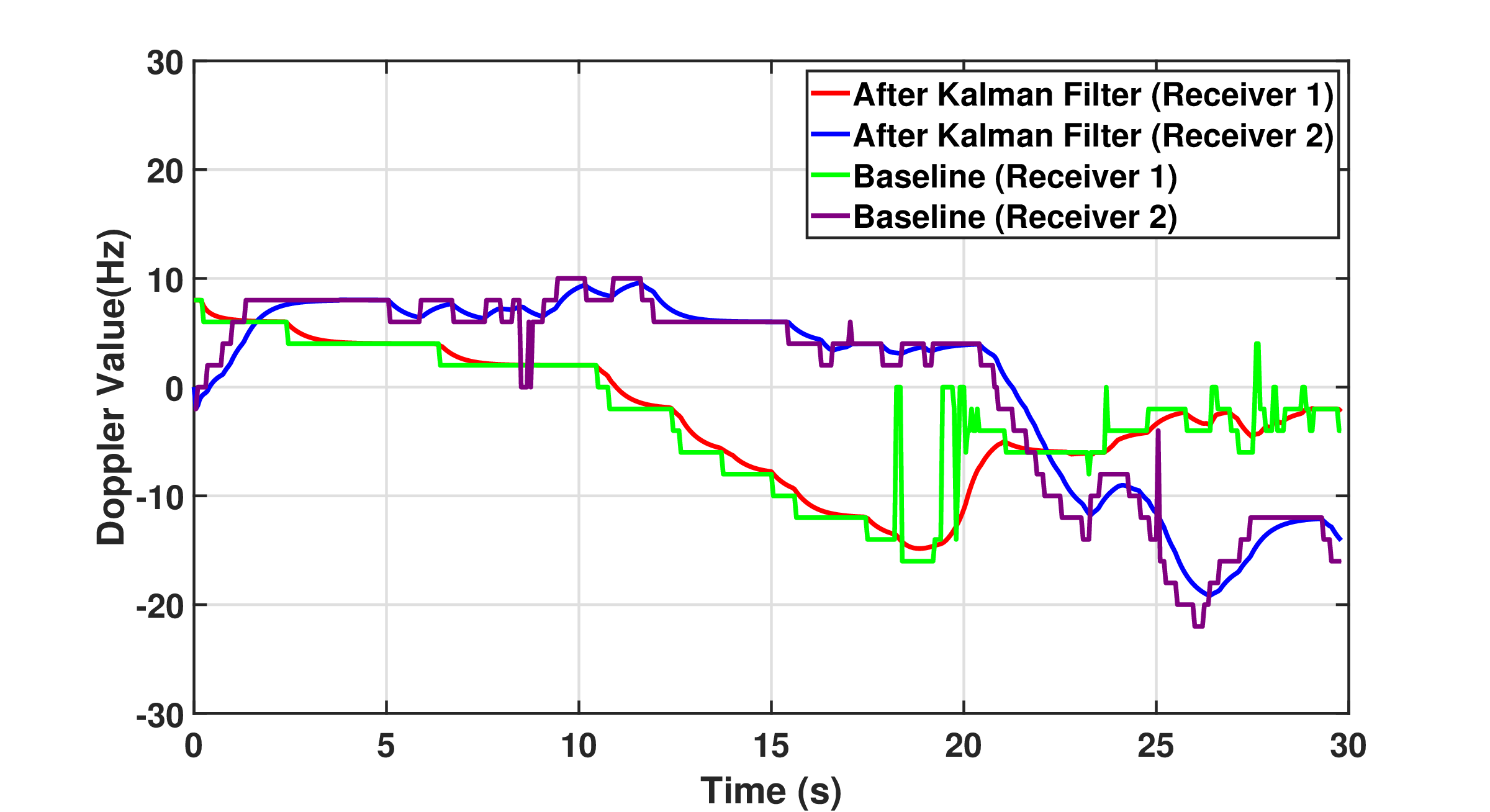}
     \vspace{-1mm}
    \caption{Detected bistatic Doppler frequency versus time with proposed and baseline schemes.}
    \label{fig6}
    \vspace{-0.3cm}
\end{figure}

Correspondingly, the estimations of bistatic Doppler frequencies versus time after Kalman filter (as proposed in Section \ref{sec3B}), as well as the baseline scheme, are illustrated in Fig. \ref{fig6}. In the baseline scheme, we select the Doppler frequency with the maximum peak value of the CAF as the UAV’s bistatic Doppler estimation. The jumps in the curves of baseline scheme are due to the false alarm.
It can be observed that the impact of the false alarms can be eliminated by the proposed algorithms in Section \ref{sec3B}, compared to the baseline scheme.

\begin{figure}[!htbp]
    \centering
    \vspace{-0.8cm}
    \subfloat[U-shaped]{
        \includegraphics[width=0.23\textwidth]{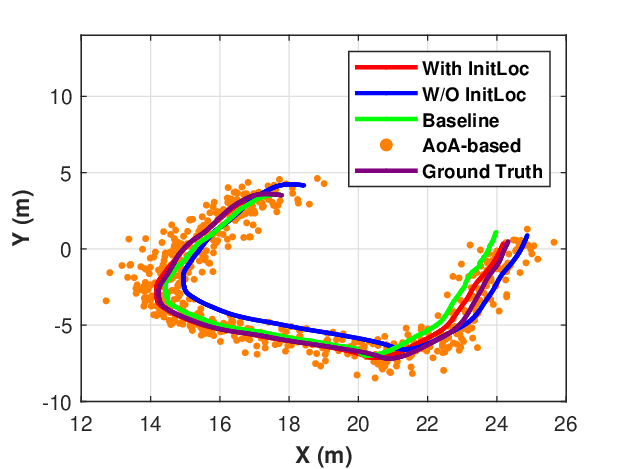}
    }
    \subfloat[Triangle]{
        \includegraphics[width=0.23\textwidth]{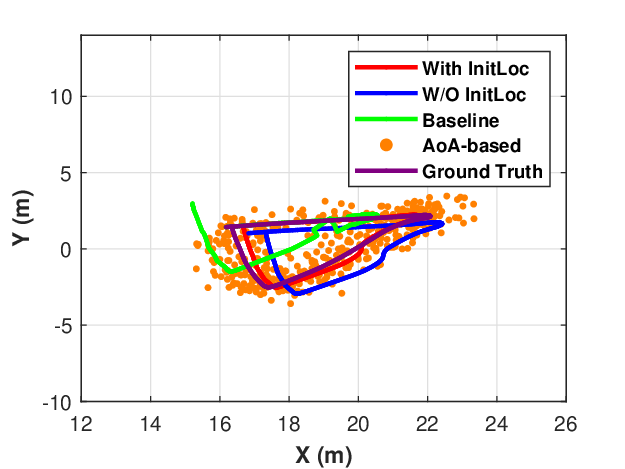}
    }
    \caption{Illustration of UAV trajectories' reconstruction.}
    \label{fig7}
    \vspace{-0.7cm}
\end{figure}

The reconstructed trajectories and the ground truths are compared in Fig. \ref{fig7}.
In the reconstruction, four scenarios are considered: (1) the UAV's initial location is perfectly known; (2) the initial location is detected with error\footnote{In this case, it is assumed that the initial UAV location is detected via the AoAs of both receivers. According to our previous work \cite{sun2024experimentalstudypassiveuav}, the error of the AoA detection with digital antenna array is typically $1.55^\circ$, which is considered in the initial location estimation of this letter.};
(3) the baseline scheme with perfect knowledge on the UAV's initial location; (4) the AoA-based localization method utilizing triangulation. In the second scenario, the initial location errors for the U-shaped and triangle trajectories are $0.9$m and $0.8$m, respectively. The performance of the last scenario is generated via simulation on true dataset, where an independent Gaussian AoA estimation noise with a mean of $0^\circ$ and a variance of $2.4025^\circ$ derived from \cite{sun2024experimentalstudypassiveuav} is added to each actual AoA.
It can be observed that the reconstructed trajectories closely align with the ground truths, as long as the initial location is perfectly known.
On the other hand, when the initial UAV location is estimated with error, the shapes of the UAV's trajectories can still be detected. This demonstrates the accuracy of the proposed tracking method.
However, the recovered triangle trajectory according to the baseline scheme clearly deviates from the ground truth. Hence, this justifies the necessity of the filtering algorithms proposed in Section \ref{sec3B}. Finally, for the AoA-based localization method, due to the Gaussian distribution of the localization errors, the reconstructed trajectory points for both U-shaped and triangle trajectories are distributed evenly on both sides of their ground truth trajectories, respectively.

\begin{figure}[!htbp]
    \centering
    \vspace{-0.5cm}
    \subfloat[U-shaped]{
        \includegraphics[width=0.23\textwidth]{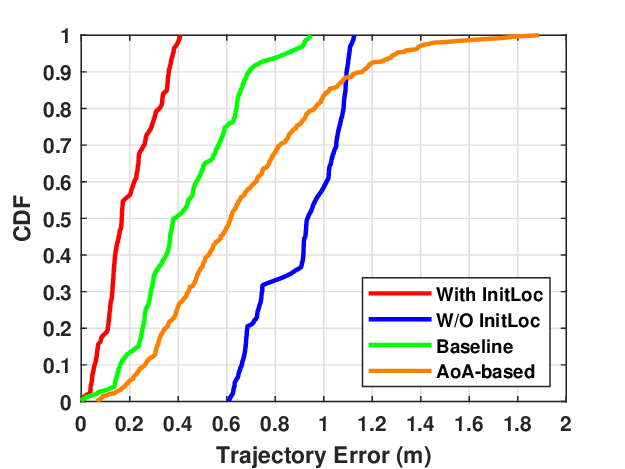}
    }
    \subfloat[Triangle]{
        \includegraphics[width=0.23\textwidth]{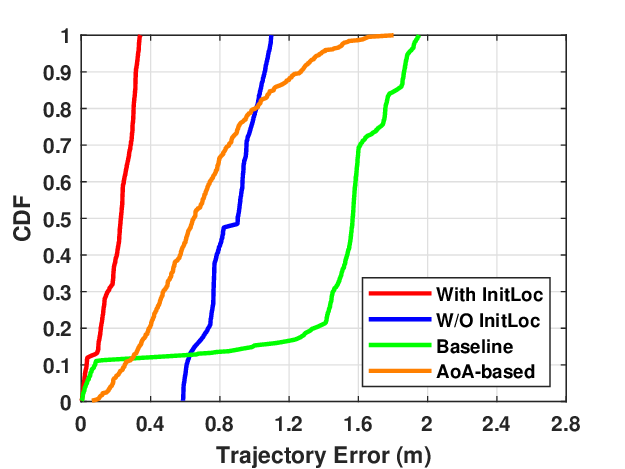}
    }
    \caption{The CDFs of the UAV tracking error.}
    \label{fig8}
    \vspace{-0.2cm}
\end{figure}

Moreover, the cumulative distribution functions (CDFs) of the tracking errors for the above trajectories are illustrated in Fig. \ref{fig8}. It can be observed that $90\%$ of the tracking errors are below $1.1$m and $0.5$m with and without initial location error, respectively. On the other hand, the tracking errors of the triangle trajectory according to the baseline scheme are significant, demonstrating that its performance is not stable. Furthermore, $90\%$ of the tracking errors in the AoA-based localization method for U-shaped and triangle trajectories are below $1.2$m and $1.3$m, respectively, which are far greater than the $0.5$m of these two trajectories demonstrated in our proposed system.

Finally, it is necessary to mention that the performance of trajectory tracking also depends on the RCS of the sensing target. The quadcopter with different payloads would lead to different RCSs, and hence, different distributions of tracking errors. It is shown by FEKO simulation that with the same payload volume, the cylinder and sphere payloads would lead to around $7$dB and $16$dB RCS degradation compared with the current cuboid payload. Moreover, $22$dB RCS degradation is shown by simulation if there is no payload. 

\vspace{-0.4cm}
\section{Conclusions} \label{sec6}
In this letter, a dual-bistatic sensing system for UAV trajectory tracking via downlink LTE signals is proposed and demonstrated. The system consists of two receivers, which detect the bistatic Doppler frequencies of the target UAV at different directions. The experimental results demonstrate that when the target UAV is around $200$m away from the LTE eNBs and $30$m from the sensing receivers, $90\%$ of the tracking errors are less than $0.5$m. Since there is no signal emission, the sensing system might be densely deployed on the top of buildings in urban area for fine-grained monitoring of UAVs.
\vspace{-0.2cm}
\bibliographystyle{IEEEtran}
\bibliography{IEEEabrv,Reference}
\end{document}